\documentclass[cits]{PoS}

\usepackage{wrapfig}

\title{Nucleosynthesis simulations for a wide range of nuclear production sites from NuGrid}

\ShortTitle{Nucleosynthesis simulations}

\author{\speaker{F. Herwig}$^{ab}$,
M. E. Bennett$^{b}$,
S. Diehl$^{c}$,
C. L. Fryer$^{d}$,
R. Hirschi$^{be}$,
A. Hungerford$^{d}$,
G. Magkotsios$^{fg}$,
M. Pignatari$^{bf}$,
G. Rockefeller$^{d}$,
F. X. Timmes$^{g}$,
and P. Young$^{g}$\\
\llap{$^a$}Dept. of Physics \& Astronomy, Victoria, BC, V8W 3P6, Canada\\
\llap{$^b$}Astrophysics Group, Keele University, ST5 5BG, UK\\
\llap{$^c$}Theoretical Astrophysics Group (T-6), Los Alamos National Laboratory, Los Alamos, NM, 87544, USA\\
\llap{$^d$}Computational Methods (CCS-2), Los Alamos National Laboratory, Los Alamos, NM, 87544, USA\\
\llap{$^e$}IPMU, University of Tokyo, Kashiwa, Chiba 277-8582, Japan\\
\llap{$^f$}Joint Institute for Nuclear Astrophysics, University of Notre Dame, IN, 46556, USA\\
\llap{$^g$}School of Earth and Space Exploration, Arizona State University, Tempe, AZ 85287, USA\\
E-mail:\email{fherwig@uvic.ca}
}

      \abstract{
  Simulations of nucleosynthesis in astrophysical environments are at
  the intersection of nuclear physics reaction rate research  and
  astrophysical applications, for example in the area of galactic
  chemical evolution or near-field cosmology. Unfortunately, at
  present the available yields for such applications are based on
  heterogeneous assumptions between the various contributing nuclear
  production sites, both in terms of modeling the thermodynamic
  environment itself as well as the choice of specifc nuclear reaction
  rates and compilations. On the other side, new nuclear reaction rate
  determinations are often taking a long time to be included in
  astrophysical applications. The NuGrid project addresses these
  issues by providing a set of codes and a framework in which these
  codes interact. In this contribution we describe the motivation,
  goals and first results of the NuGrid project. At the core is a new
  and evolving post-processing nuclesoynthesis code (PPN) that can
  follow quiescent and explosive nucleosynthesis following multi-zone
  1D-stellar evolution as well as multi-zone hydrodynamic input,
  including explosions. First results are available in the areas of AGB
  and massive stars.}

\FullConference{10th Symposium on Nuclei in the Cosmos\\
		 July 27 - August 1 2008\\
		 Mackinac Island, Michigan, USA}

\begin{document}

\section{Introduction}
Nuclear astrophysics combines nuclear physics of astrophysical
relevance with the simulation of nuclear production sites in stellar
evolution and explosions, and ultimately with abundance observations
in stars and galaxies and measurements in pre-solar grains. Numerous
compilations of yield data, for applications such as chemical
evolution of galaxies, have been presented, based on different
modeling assumptions. For example, in the area of massive stars the
compilation of Woosley \& Weaver (1995) \nocite{woosley:95} provides
yields which is based on solving a nuclear network together and in
lock-step with the stellar evolution code. A similar approach is
applied for AGB star yields \cite{herwig:04a,karakas:07}, although
yields based on synthetic models are also still in use
\cite{marigo:01b}. The latter was justified by the significant labour
involved in full stellar evolution tracks of the advanced phases of
stellar evolution, where most of the interesting nucleosynthesis takes
place. It is largely for this reason that we still don't have yield
tables that cover low-mass \emph{and} massive stars (including
explosive yields) for a meaningful range of metallicities and both
light \emph{and} heavy elements with internally consistent physics
assumptions, including the nuclear physics data.  However, such
comprehensive yield data is required, for example in near-field
cosmology applications \cite{venn:04,font:06a}.

In addition, new results, for example on the hydroynamic nature of
convective boundary mixing (Woodward et\,al., this vol.\ and
\cite{meakin:07b}), need to be included in new yield calculations as
quickly as possible to make them available for comparison with
observations. Finally, the nuclear physics community needs to prioritze their
efforts through the ability to run numerical nucleosynthesis
experiments in realistic stellar production environments. On the
other hand,

In order to address these issues we have pooled capabilities and
expertise in the nucleosynthesis grid (NuGrid) collaboration to create
a new simulation library and nucleosynthesis code capability. In this
paper we describe our approach and report first results, for example
from AGB s~process and massive star nucleosynthesis. Other results
relating to the NuGrid project have been presented at this conference
by Hirschi et\,al., Pignatari et\,al., Fryer et\,al.,  Diehl et\,al.,
Hungerford et\,al. and Rockefeller et\,al.\, .

\section{NuGrid nucleosynthesis post-processing}

We have developed a new post-processing nucleosynthesis (PPN) code,
and we are developing a stellar evolution and explosion (SEE) database,
including an interface that allows these two components to communicate
efficiently. The ultimate goal is to combine these two tools to create
the needed comprehensive and internally consistent yields tables.
The design goals of the PPN codes are (1) a capability of post-processing a
wide range of thermodynamic environments from both 1D stellar
evolution codes and trajectories from hydro-simulations (explosions
and stellar hydro, both grid and particle), (2) comprehensive yet
flexible nuclear physics input, and (3) to match resolution, detail,
accuracy and precision of TD simulations, observations and nuclear
physics.

The core unit of the PPN code implementation is a nuclear network
kernel that consists of a physics package and a solver package. The
nuclear network kernel evolves one nuclear network cell over one time
step. There are three drivers that use the same network kernel. The
single-zone driver (SPPN) is used for simple one-zone network
experiments with either analytic, algorthmic or tabulated
thermodynamic input. The multi-zone driver (MPPN) post-processes the
output of one-dimensional spherically symmetric stellar evolution
codes, while the trajectory driver (TPPN) deals with trace particle
data from hydrodynamic simulations.

\begin{wrapfigure}{p}{0.5\textwidth}
\centering
   \includegraphics[width=0.5\textwidth]{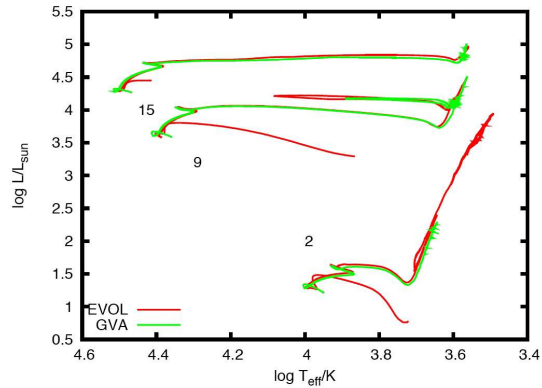} 
   \caption{Stellar evolution tracks from EVOL and GVA (Geneva code)
     for different masses as indicated, without rotation, identical
     initial composition and an overshoot parameter
     ($f_\mathrm{ov}=0.014$ for EVOL and $\alpha_\mathrm{ov}=0.2$ for
     GVA) that has been choosen so that the width and duration of the
     main-sequence match.}
   \label{fig:tracks}
\end{wrapfigure}
The nuclear physics package includes most major nuclear data
compilations, i.e.\ Basel reaclib \cite{thielemann:87}, Kadonis
\cite{dillmann:06}, NACRE \cite{angulo:99}, Illiadis et\,al. (2001,
\cite{iliadis:01}), Cauhglan \& Fowler (1988, \cite{caughlan:88}), the
nuclear data online interface Bruslib \cite{aikawa:05}, Oda
et\,al. (1994, \cite{oda:94}) and Fuller et\,al.\ (1985,
\cite{fuller:85}). The network includes NSE with T-dependent partition
functions and mass excesses from reaclib and Coulomb screening from
Calder et\,al. (2007, \cite{calder:07}). The network is dynamically
built in two steps. In a first configuration step the master set of
isotopes out of a maximum of 5180 is selected using simple
configuration instructions. Based on this master set the actual
network is adjusted dynamically in size for each network kernel calculation,
so that the solution in every network cell is performed for the
optimal selection of isotopes. The solver package relies at this time
on a Newton-Raphson, fully implicit implementation with full precision
control and adaptive sub-time stepping. We are also implementing a
variable order method for improved accuracy \cite{timmes:99}. The
multi-zone and trajectory drivers are parallelized through a simple
master-slave strategy, implemented in the distributed memory standard
MPI. The parallel MPPNP driver has been run on up to 150 cores,
although in most practical applications we are running post-processing
grids of $\sim 250$ shells on 60-80 cores. The MPPNP drivers provides
three grid options: static, input grid or adaptive grid.

The interface to the stellar evolution and explosion data is defined
through the custom library USEEPP\footnote{USEEPP = Unified Stellar
  Evolution and Explosion Post-Processing} built on top of the
platform independent hdf5\footnote{http://www.hdfgroup.org/HDF5}
standard. The SEE database is populated with low-mass tracks with the
EVOL code (Pignatari et\,al., this vol.) and with tracks from the Geneva
code (Hirschi et\,al., this vol.) and the Tycho code \cite{young:05a}
for massive stars. Within the physics options implemented in these
codes we are calibrating the free parameters to obtain the largest
possible internal consistency (Fig.\ \ref{fig:tracks}). Explosion
simulations in the SEE database are provided as described in Fryer
et\,al.\ (this vol.).

\begin{figure}[tbp]
   \includegraphics[width=0.5\textwidth]{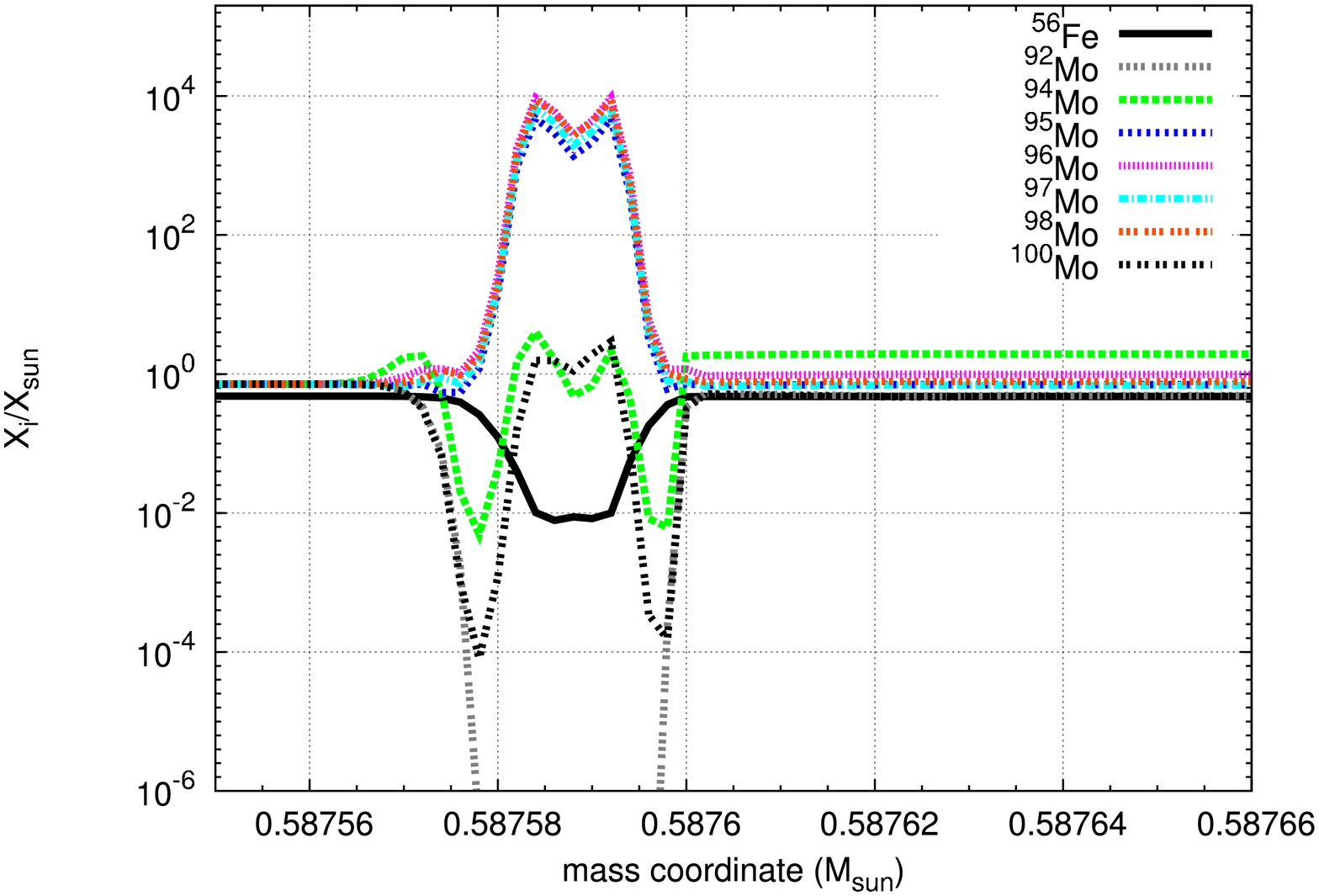}    
   \includegraphics[width=0.5\textwidth]{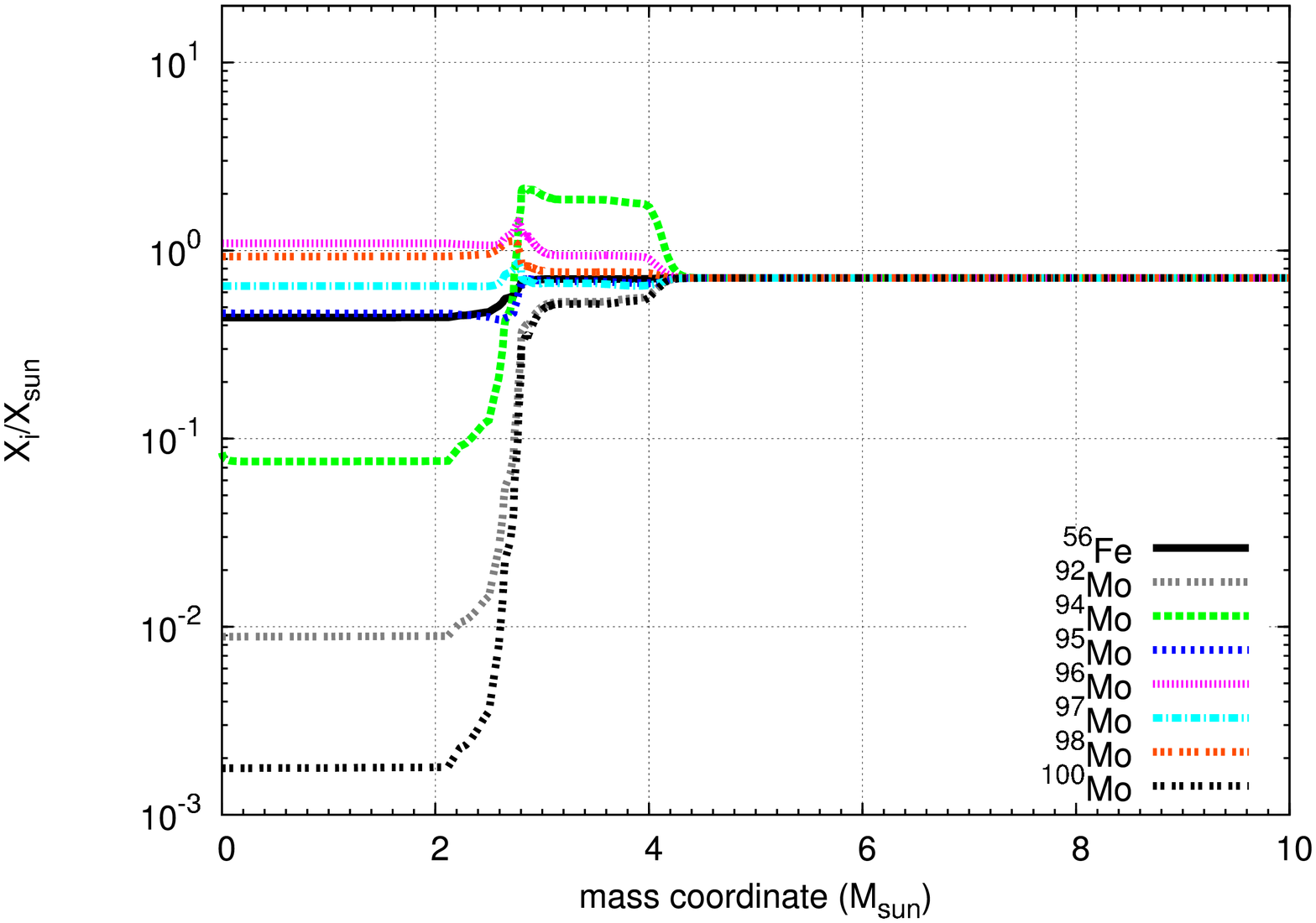} %
   \caption{Post-processing results for the Mo isotopes and the
     s-process seed $^{56}$Fe at the end of the interpulse
     $^{13}C$-pocket nucleosynthesis (left) as in Pignatari et
     al.\ (this vol.) and in He-core burning in a 15M$_\odot$ model
     (right) as described in more detail in Hirschi et\,al.\ (this vol.)}
   \label{fig:spr}
\end{figure}
One potential problem with the post-processing approach is the need to
exactly match nuclear reactions that produce the majority of the
nuclear energy in the stellar evolution and post-processing code.  We
have a special interface in the physics package to either reproduce
exactly the same reaction rate source as used in the stellar evolution
code, or to introduce the same subroutine with the same interpolation
or fit formula evaluation algorithm as used in the stellar evolution
calculation. In addition, we save up to ten control abundances that
are primarily linked to the energy production in the USEEPP format,
and monitor any differences that may develop between the original and
post-processing nucleosynthesis.  Another problem may be to correctly
map the original stellar evolution or explosion calculation into the
post-processing code. Through our USEEPP IO library all
thermodynamic and mixing data in all mass zones and all time steps of
all tracks are saved, which takes on average 5GB per full stellar evolution
track. We are then able to either post-process on the original stellar
evolution grid, or accomodate special grid requirements of the
nucleosynthesis.  Thus, our post-processing approach is accurate with
the added benefit of additional grid options, updated nuclear physics
and optionally higher-order solvers.  The results are as reliable as yields
calculated with an extra nucleosynthesis step inlined into the
stellar evolution simulations, with a larger network than used for the
energy generation feeding into the stellar structure solver
\cite{woosley:95,karakas:07}.  Contrary to the latter approach we can
rerun our post-processing with any nuclear phyiscs input at minimal
human labour cost. This method is affordable, with a full post-process
run of one stellar evolution track sequence ($10^5$ time steps, $\sim
300$ pp grid zones) taking 2 days on $\sim 60 \dots 80$ cores.

We have so far populated the SEE database with both low-mass and
massive star tracks (Fig.\, \ref{fig:tracks}). A major advantage
of our approach is the ability to calculate the nucleosynthesis in
both regimes with the same MPPNP code and the same nuclear physics
data. At solar-like metal content s-process contributions to Mo come
from AGB stars. However, as discussed in this volume by Hirschi et
al., models of very low metal content with rotation may produce
significant amounts of s-process elements between Sr and Ba, including
Mo. Fig.,\ \ref{fig:spr} demonstrates by example of Mo nucleosynthesis
in an AGB and a massive star environment how we will use our NuGrid
to investigate nucleosynthesis in a comprehensive and consistent way.

\acknowledgments

FH was supported 
Marie Curie International Reintegration Grant within the
6$^\mathrm{th}$ European Community Framework Programme, grant
MIRG-CT-2006-046520. MP ackowledges support through NSF grants PHY
02-16783 (Joint Institute for Nuclear Astrophysics).

\bibliography{astro}

\end{document}